# COMPENSATION OF BEAM–BEAM AND SPACE-CHARGE EFFECTS: EXPERIENCE TO-DATE AND NEAR-FUTURE OPPORTUNITIES *

V. Shiltsev[#], FNAL, Batavia, IL 60510, USA

*Abstract*

Beam-beam interactions and space-charge effects belong to the category of the most long-standing issues in beam physics, and even today, after several decades of very active exploration and development of counter-measures, they still pose the most profound limitations on performance of accelerator facilities. In this brief review we consider past experience in active compensation of these effects and possible new schemes for further exploration in near-future, in particular, within the framework of the electron-ion collider R&D.

## BEAM-BEAM COMPENSATION

Significant experimental advances on beam-beam compensation have been made since late 1990's. The perturbative or disruptive effect of the beam-beam interactions is a basic limitation to increasing the luminosity of colliders and is a strong incentive to devise and study compensation methods. With the advent of hadron colliders with a small bunch spacing, it becomes necessary to consider the compensation of both the head-on and the long-range beam-beam effects. By nature, the head-on beam-beam force derives from a Poissonian potential while the magnetic force of optical lenses is Laplacian, defeating attempts at correcting one by the other, at least exactly. The long-range beam-beam effect is however close to Laplacian for realistic beam-beam separations, opening new compensation possibilities.

**Compensation of the head-on beam-beam effect.** *Four-beam compensation* - If four beams are made to collide at the same point, with, for each direction of propagation, one beam of particles and one beam of antiparticles of equal intensity and transverse beam sizes, there is no net electromagnetic beam-beam force. This concept giving exact compensation with a potential of substantially improved performance [1] was experimented in DCI at Orsay, France ($e+/e-$ at 0.8 GeV cme) – see Fig.1. While in a three-beam weak-strong configuration, an increase by a factor of 5 of the beam-beam limit was observed, no improvement of performance was obtained in the four beam configuration [2]. Unexpected excitation of non-linear beam-beam resonances was noticed, as well as coherent signals. These observations seem in qualitative agreement with the prediction [3] that the coherent beam-beam limit is not improved by the four-beam system, due to the cancellation of the beam-beam driven Landau damping. This coherent limit is however expected at somewhat higher beam-beam parameter than observed [4]. A four beam compensation concept has been contemplated for $e+/e-$ linear colliders but found to be plagued by plasma instabilities, which lead to significant charge separation and luminosity reduction even at very small initial bunch displacement errors [5, 6].

*Electron-lens compensation* - The compensation of the beam-beam effect by an auxiliary beam is a variation of the above solution that allows a drastic simplification appropriate for high-energy colliders, however with some limitations. An auxiliary electron beam of low energy is prepared in a source, made to collide with the main beam in a strong solenoidal field and dumped after the interaction, suppressing the possibility of coherent coupling, suspected to have plagued the four-beam concept. The auxiliary beam shall have suitable charge/direction of propagation for compensation and the same transverse positions and sizes as the perturbing beam [7]. It should ideally be positioned at the interaction point. This is however not possible in practice and mitigations must be carried out in the compensation strategy. The first study and implementation of an electron lens was done at the Tevatron (see Fig.2, with a schematic view of the lens).

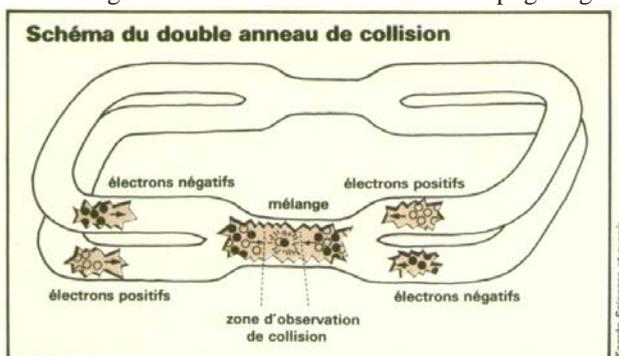

Figure 1: Scheme of four beam collisions ($e+e-e+e-$) in the DCI storage ring (see text).

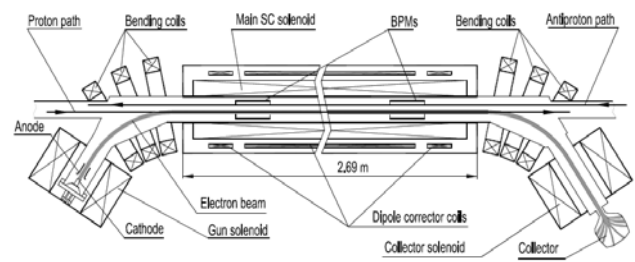

Figure 2: Layout of the Tevatron Electron Lens (TEL).

Comprehensive tests [8] have led to the optimization of the transverse electron current density for linear corrections (constant electron density over the main beam extent and smooth tails) and have demonstrated successfully the compatibility of electron lenses with the operation of superconducting hadron colliders and proton/antiproton tuneshift induced by electrons as high as $dQ$=0.01. It was shown that the noise of the electron current can be reduced to a level that does not cause emittance blow-up of the main beam. This complex instrument reached a high reliability. It has been used to correct the primary beam-beam limitation of the Tevatron performance related to a tune spread along the bunch train. By using the lenses as pulsed bunch-by-bunch focusing elements, the tunes of the normal and Pacman bunches can be equalized, leading to a noticeable gain in lifetime (Fig.3).

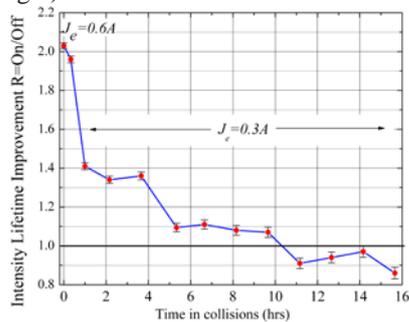

Figure 3: Intensity lifetime of proton bunch 12 when the TEL is consequently switched off and on during the Tevatron high-energy physics collision store [8].

Studies of nonlinear beam-beam correction have been undertaken in the Tevatron during dedicated machine development studies at the end of the Collider Run I. Electron beam from a Gaussian e-gun of the TEL-2 was aligned on the antiproton bunches. The conditions of the studies were not perfect (proton beam was too broad and the beam‑beam effect of protons on antiprotons was too small) and indications of the beam lifetime improvement were not conclusive [9]. In very near future, the head-on beam-beam compensation will be tested in RHIC: two e-lenses with Gaussian e-guns have been installed in RHIC's IR10 – one for each proton beam – and a factor of 2 improvement in the beam-beam parameter and, correspondingly, in the luminosity gain are expected [10].

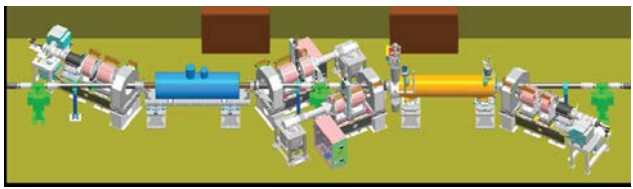

Figure 4: Layout of the two electron lenses in RHIC's IR10. There are three beams in each lens - the two proton beams and the electron beam acting on one of the proton beams. The proton beams are vertically separated.

*Compensation by octupoles* - In VEPP4 (*e+/e-* at 5.3 GeV), the compensation of the cubic beam-beam non-linearity using octupoles was systematically investigated by scanning the tune plane and the lattice octupole strength [11]. By measuring the beam loss rate, it was demonstrated that the width of resonant islands could be increased or decreased depending on the combination of beam-beam effect and powering of the lattice octupoles. In operating conditions, however, no clear improvement could be obtained in VEPP-4, neither in CESR, where similar trials were made.

*Plasma Suppression of Beam-Beam Interaction* - The influence of one beam on the other can be compensated if the collisions take place in dense enough plasma. A good compensation requires an overdense plasma $n_{beam}/n_{plasma} \ll 1$, a short plasma period and a small current skin depth. Introduction of the plasma in the interaction region gives rise to parasitic collisions of the beam particles with the plasma ions and electrons which cause a growth of the beam emittance, particle losses and higher detector background. It is shown in [12] that in TeV muon colliders, the overdense plasma with $n_{plasma} \sim O(10^{19}$ cm$^{-3}$) can easily suppress the beam-beam tune-shift parameter several times without degrading the beam lifetimes. Beamstrahlung effects in linear e+e- colliders can be suppressed too, if the required very high densities of plasmas of the order of or exceeding the electron densities in solids can practically be achieved [13].

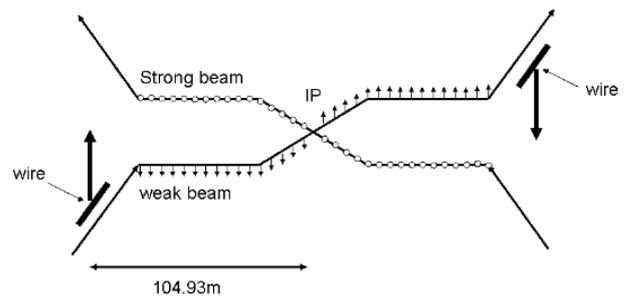

Figure 5: Principle of wire compensation of long-range beam-beam effects.

**Compensation of long-range beam-beam effects.** *Wire compensation* - The principle of an almost exact compensation of the long-range beam-beam effect with wires was proposed in [14] for the LHC, where the long-range beam-beam interactions are clustered on either side of the interaction points. Each cluster is compensated by one wire positioned with the following requirements: the smallest betatron phase advance between clustered perturbations and compensation (a few degrees); same transverse beam aspect ratio as at the long-range perturbations; beam-wire separation identical to the beam-beam separation at the long-range interaction points when expressed in units of the rms transverse size; the integrated wire current equal to the sum of the integrated beam currents in the cluster; sufficient separation between the beam channels to install the movable wire set-up - see Fig.5.

Extensive numerical investigations and verifications of the compensation efficiency and robustness have been carried out, wire excitation and compensation units were fabricated, installed and studied in the SPS [15] and RHIC [16]. Compensation proper has only been investigated in the SPS, where one wire unit simulates the strong beam and another, installed at a betatron phase shift modelling the LHC compensation scheme, compensates the perturbations of the former.

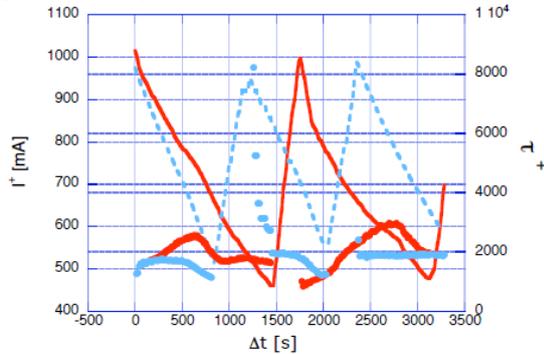

Figure 6: DAFNE e+ beam current and lifetime as a function of time: wires on (red) and wires off (cyan).

If the long-range beam-beam interactions are not clustered but distributed all around the machine, wire compensation meets the same difficulties as non-local non-linear corrections. It was simulated for the Tevatron and shown to yield a small improvement of stability if not a worsening [17]. On the contrary, for DAFNE (INFN, Fracsati), a non-local wire correction was studied and implemented, leading to a significant increase of the beam and luminosity lifetimes– e.g., during the machine studies it was shown that is possible to improve the lifetime $\tau+$ of the 'weak' positron beam in collision ~30% and deliver the same integrated luminosity with less injections [18] – see Fig. 6.

**Outlook for active compensation of beam-beam effects.** The most actively, the beam-beam compensation is currently being pursued by the RHIC team [19] and as soon as in 2015-2016 one may expect the first results of initial head-on compensation studies. Another active development is taking place at Fermilab (supported by US LARP) and CERN to simulate, design and install either current-carrying wires or electron lenses (several modifications) to compensate long-range and possibly head-on beam-beam interaction effects in the high-luminosity LHC [20, 21, 22] which is currently scheduled to start around 2023.

## SPACE-CHARGE COMPENSATION

The effects of the space-charge forces are seriously limiting performance of high intensity proton accelerators [23]. While most of the following discussion is devoted to the effects of transverse space-charge (SC) forces, here we start with bringing attention to the longitudinal SC fields which can generate substantial distortion of the rf-generated potential wells, fill the extraction kicker gap in the beam, affect the incoherent synchrotron tune spread, and have the potential for causing instability and longitudinal emittance growth. The net effective voltage per turn resulting from the space-charge self voltage and the ring inductive wall impedance $\omega_0 L$ is proportional to the slope of the beam current distribution $e\beta c\, \lambda(s)$ and can be expressed as:

$$V_s = \frac{\partial \lambda(s)}{\partial s}[\frac{g_0 Z_0}{2\beta\gamma^2} - \omega_0 L]e\beta cR$$

where $R=c/\omega_0$ is the average machine radius, $Z_0=377$ Ohm and $g_0=1+2ln(b/a)$ is the geometric space-charge constant, $a$ and $b$ are the beam radii and vacuum-chamber aperture. By introduction a tunable inductance $L$, e.g. of ferrite rings, the term in brackets and, consequently, the space-charge effect may be substantially reduced or cancelled at some chosen energy [24].

This concept has been experimentally proven at the LANL Proton Storage Ring at LANL where three inductive inserts, each consisting of 30 "cores" of a cylindrically shaped ferrite with thickness of 1 inch, inner diameter of 5 inches, and an outer diameter of 8 inches, were installed. The magnetic permeability of the ferrite could be adjusted by introducing current into solenoids wound around the ferrite so that in the MHz range of frequencies the longitudinal space charge impedance of the machine was compensated [25]. A strong longitudinal instability was noticed at much higher frequencies of about 75 MHz, but it was later suppressed by heating the ferrite to a temperature of 130∘C to make it more lossy. The inserts have proven beneficial in raising the threshold for the two-stream electron-proton (*e-p*) instability at PSR – see Fig.7 - and achieving shorter bunch length.

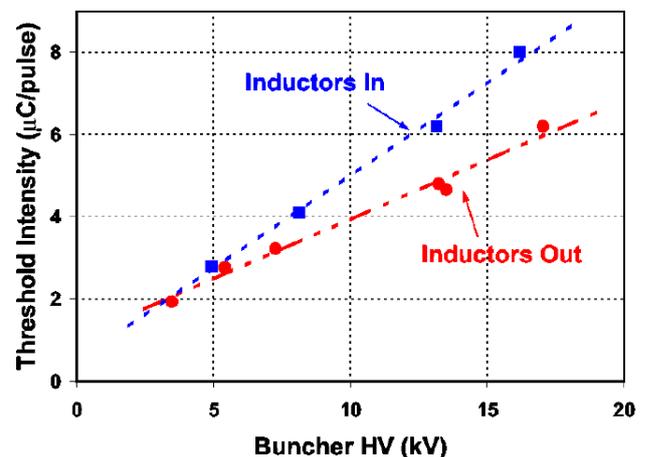

Figure 7: The PSR electron-proton instability threshold vs rf buncher voltage with and without the inductive inserts.

**Transverse Space-Charge Compensation**

There are several methods to compensate transverse SC effects which often manifest themselves in the form of beam loss, core emittance growth, and halo formation.

In the case of **passive neutralization,** the SC force of a proton beam can compensated by ionization electrons, electron cloud, or negative ions which are approximately at rest longitudinally, but move transversely during the beam passage. The compensation condition reads $\eta \approx 1/\gamma_p^2$. Neutralized low energy beams of heavy ions have successfully transported in a number of linear accelerators [26]. A factor of 9.5 increase of the maximum circulating beam current above (coherent) SC limit was achieved at the Novosibirsk 1 MeV proton ring by increasing the residual gas pressure in excess of $10^{-4}$ Torr and accumulation of ionization electrons [27]. The beam lifetime was very short and transverse and longitudinal proton distributions not well controlled.

Optimum compensation requires that the transverse electron and beam distributions are matched. That could be achieved by confining the electrons transversely with strong solenoid fields to "columns" and using electrostatic electrodes to fine tune the charge density – see Fig.8 from [28]. Strong magnetic field also stabilizes electron "column" motion and prevents coherent e-p instability. Simulations show significant reduction of SC induced emittance growth with only few "columns" occupying a small fraction $\eta$ of the ring circumference [28].

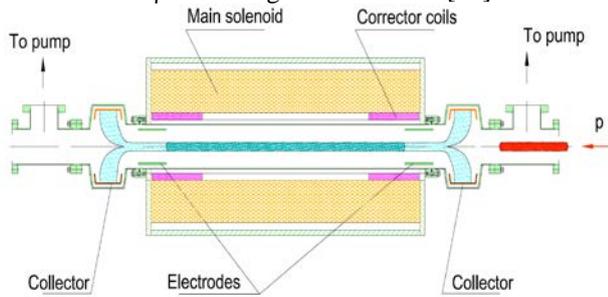

Figure 8: Schematic layout of an "electron column" for space-charge compensation.

It is to be noted that the process of accumulation of ionization electrons is strongly dependent on presence of ions. In the case of fast transiting ions the growth rate of diocotron modes is relatively small and drops strongly with magnetic field strength. In the case of slow trapped ions the growth rate of diocotron modes is defined by the neutralization (space-charge compensation) level solely, and thus may be very dangerous. Fig.9 shows results of the simulation of the plasma formation and trapping in the e-column using WARP 3D code [29]. There are various stabilization and damping techniques, out of which the most effective has to be chosen according to plasma and trap parameters. For example, a "rotating wall" technique might be used to compensate the radial transport caused by the mode damping processes

**Electron lenses**, in which externally generated electron beam with matched transverse distribution collides with the proton beam inside a strong solenoid field, could also compensate the SC tune shift [30]. Assuming the total length of the lenses $L$, distributed around the ring, and an electron beams co-propagating with the proton beam, the electron current needed per lens is [31] $J_e = (B_f \kappa e c N_p/L) \beta_e/( \gamma_p^2 (1 - \beta_p \beta_e))$, and for many accelerators of interest lays in the range of 1-10A for 10-40 keV electrons (here $\kappa$ denotes the degree of compensation, $B_f$ – proton bunching factor). These parameters are close to those of the operating Tevatron electron lenses. The SCC by lenses works better if the electron current is modulated to match longitudinal profile of proton bunches [32]. A practical method to achieve necessary time modulation of the electron focusing forces has been recently proposed in [33].

**Passive cancellation** of the next-to-leading term in the s.-c. force is possible by **octupole** fields. For a round beam, the 4th order of term of the direct s.-c. potential varies as $(x^4 + 2x^2y^2 + y^4)$, while the potential of an octupole is proportional to $(x^4 - 6x^2y^2 + y^4)$. Therefore, at least two families of octupoles are needed to reduce the SC tune spread, which are placed at locations with either peak and intermediate values of the beta function, respectively. The beta functions should sufficiently vary over the length of an optical cell, e.g., by a factor 2 or more.

**Pole-face windings** allow precise adjustments of the tune shift with transverse position up to a high order. At the CERN ISR, 24 pole-face windings modifying the local magnetic field were used to correct the horizontal and vertical indirect SC tune shift plus the next 4 orders in their Taylor expansions with respect to the horizontal position. The correction increased the maximum ISR beam current 15 times [34].

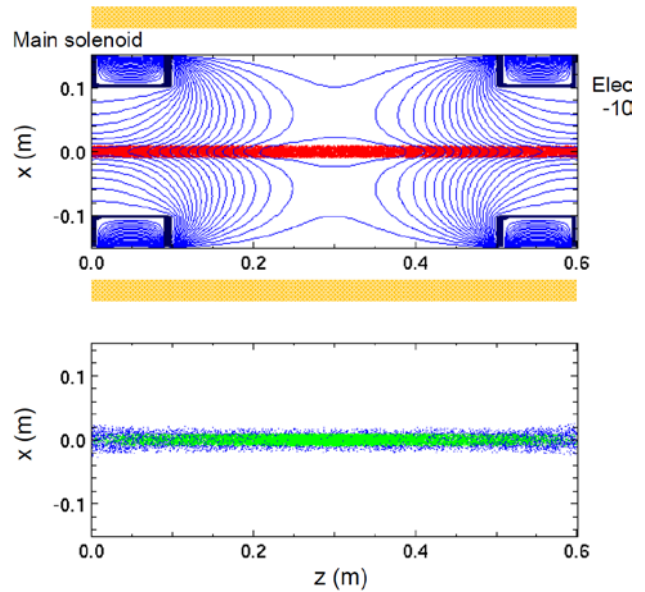

Figure 9: WARP 3D simulation of the plasma formation and trapping in the e-column. A uniform solenoid keeps both electrons (green dots) and ions (blue dots) in the beam path. Ions are slowly escaping longitudinally [29].

Recently proposed **fully nonlinear but integrable lattice** accelerators have promise to accommodate extraordinary large tune spreads in circulating beams without driving losses (resonance free optics) [35]. Comprehensive numerical studies of SC dynamics in the Integrable Optics rings have been started.

## OUTLOOK: EXPERIMENTS AT ASTA

There are many challenges for the proposed space-charge compensation methods which call for experimental verification, including stability of the electron-proton system (transverse motion), (dynamic) matching of transverse p-charge distribution, appropriate longitudinal compensation (for not-flat proton bunches), (dis)advantages of electron lenses vs electron columns, technology and practical implementation (in existing facilities), etc. A unique chance for carrying out the much needed dedicated studies is offered by new Fermilab's Accelerator R&D facility ASTA [36].

The Advanced Superconducting Test Accelerator (ASTA) at Fermilab incorporates a superconducting radiofrequency (SRF) linac coupled to a photoinjector and small-circumference storage ring capable of storing electrons or protons. ASTA will establish a unique resource for R&D towards Energy Frontier facilities and a test-bed for SRF accelerators and high-brightness beam applications. The unique features of ASTA include: (1) a high repetition-rate, (2) one of the highest peak and average brightness within the U.S., (3) a GeV-scale beam energy, (4) an extremely stable beam, (5) the availability of SRF and high-quality beams together, and (6) a storage ring capable of supporting a broad range of ring-based advanced beam dynamics experiments [36, 37].

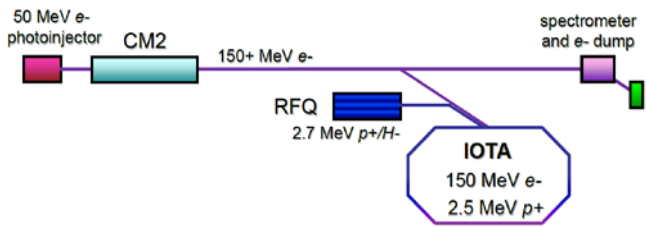

Figure 10: Layout of ASTA facility for accelerator R&D.

The experiments planned to be carry out at the ASTA's new storage ring (IOTA) include the initial set which requires well-qualified narrow electron beam: Integrable Optics test with non-linear magnets, Integrable Optics test with e-lens(es), optical stochastic cooling Test, electron quantum wavefunction size, etc; which will be followed by experiments with 2.5 MeV H- and protons: SC modes and dynamics in the ring with and without integrable optics, SC compensation with e-columns or/and e-lenses, H- halo and stripping, beam and halo diagnostics, etc.

Author greatly acknowledges fruitful discussions on the subjects of the review and input from many colleagues, including Yu.Alexahin, W. Foster, A. Burov, F. Zimmermann, V.Litvinenko, V. Kapin, Ya.Derbenev, V. Dudnikov, V. Danilov, J.P.Koutchuk, A. Valishev, M. Chung, S. Nagaitsev, G. Stancari, A. Kabantsev.